\documentclass[prl,showpacs,twocolumn,superscriptaddress,aps,a4paper]{revtex4-1}
\usepackage{dcolumn}
\usepackage{amsmath}
\usepackage{graphicx}
\usepackage{latexsym}
\usepackage{amsfonts}
\usepackage{amssymb}
\DeclareGraphicsExtensions{.pdf,.gif,.jpg}

\def\a{\alpha}
\def\b{\beta}

\def\g{\gamma}

\def\L{\Lambda}

\def\d{\delta}

\def\ve{\varepsilon}

\def\bra{\langle}
\def\ket{\rangle}

\newcommand{\be}{\begin{equation}}
\newcommand{\ee}{\end{equation}}
\newcommand{\beq}{\begin{eqnarray}}
\newcommand{\eeq}{\end{eqnarray}}

\begin{document}

\title{Advances in nonequilibrium transport with long-range interactions}

\author{E. Perfetto}
\affiliation{Dipartimento di Fisica, Universit\'a di Roma Tor Vergata, Via della Ricerca Scientifica 1, I-00133 Rome, Italy}

\author{G. Stefanucci}
\affiliation{Dipartimento di Fisica, Universit\'a di Roma Tor Vergata, Via della Ricerca Scientifica 1, I-00133 Rome, Italy}
\affiliation{INFN, Laboratori Nazionali di Frascati, Via E. Fermi 40, 00044 Frascati, Italy}
\affiliation{European Theoretical Spectroscopy Facility (ETSF)}

\author{M. Cini}
\affiliation{Dipartimento di Fisica, Universit\'a di Roma Tor
Vergata, Via della Ricerca Scientifica 1, I-00133 Rome, Italy}
\affiliation{INFN, Laboratori Nazionali di Frascati, Via E. Fermi 40, 00044 Frascati, Italy}

\begin{abstract}

The effects of long-range interactions in quantum transport 
are still largely unexplored, mainly due to the difficulty of devising 
efficient embedding schemes. 
In this work we present a substantial progress in the interacting 
resonant level model by reducing the problem to the solution of 
Kadanoff-Baym-like equations with a {\em correlated} embedding 
self-energy. The method allows us to deal with short- and long-range 
interactions and is applicable from the transient to the steady-state regime.
Furthermore, memory effects are consistently incorporated and the 
results are not plagued by negative densities or non-conservation of 
the electric charge. 
We employ the method to calculate densities and 
currents with long-range interactions appropriate to low-dimensional 
leads, and show the occurrence of a jamming effect which drastically 
reduces the screening time and suppresses the zero-bias conductance. 
None of these effects are captured by short-range dot-lead interactions.

\end{abstract}


\maketitle

Electron correlations 
have profound implications on the transport
properties of nanoscale devices\cite{book}.
{\it Local} interactions within 
small molecules or 
quantum dots contacted to leads give rise to
peculiar phenomena like Kondo effect\cite{kondo} and Coulomb 
blockade\cite{cb}, and have been the subject of several  studies.
Much less attention has been devoted instead to the {\it nonlocal} 
interactions responsible for interfacial screeening and 
polarization-induced renormalizations of the molecular levels.
%
%
Recently, a short-range (SR) dot-lead interaction has been shown 
to cause a reduction of the quasiparticle gap due to the image charge 
effect\cite{thygesen,thygesen2,bescond}. 
%
%
%
In the interacting resonant level model (IRLM) the SR  interaction
is also at the origin of a negative differential conductance with
a (interaction-dependent) power-law
\cite{boulat,nishino,karrasch,andergassen} as well as of an
overall enhancement of the off-resonance conductance  
\cite{bohr,borda}.

%

%
%

The theoretical progresses in dealing with SR dot-lead 
interactions are, unfortunately, 
not directly exportable to study long-range (LR) 
interactions, more appropriate for low-dimensional leads.
The difficulty stems from the impossibility of combining 
many-body methods with embedding techniques, hence reducing 
the problem to the evaluation of the Green's function of a finite and 
interacting open system\cite{meir,jauho}.
Recently Elste and coworkers\cite{millis} approached the problem 
using the rate equations (RE) method in the IRLM with Luttinger liquid leads.  
The RE, however, are not reliable in the transient regime 
and underestimate the steady-state polarizability of the dot, 
as we will clearly show below.
The fundamental questions which remain at present totally 
unanswered are therefore: What is the impact of a LR dot-lead 
interaction in the $I$-$V$ curve? 
How does the screening time change from SR to LR interactions?


In this Letter we consider the IRLM as the prototype model to
address the above issues. 
We study the real-time evolution of the currrent and dot-density 
after the sudden switch-on of a bias voltage for both SR and 
LR dot-lead interactions. Our results indicate that LR interactions
produce a jamming effect in the leads which
(i) shortens the screening time and (ii) 
drastically suppresses the zero-bias conductance.

%
%

The proposed methodology to conclude (i) and (ii) is 
based on a truncation of the equations of motion for dressed 
correlators. The procedure leads to  Kadanoff-Baym-like equations 
with a correlated embedding self-energy which incorporates all 
interaction and memory effects. Our approach overcomes the
{\it negative probability problem}\cite{whitney} of the RE and is, at 
the same time, charge-conserving.
The final equations are exact in the uncontacted case as well as 
in the noninteracting case and several analytic results are obtained 
in the steady-state, including a Meir-Wingreen-like formula for the 
current.
We benchmarked this formula  against recent results 
with SR interaction obtained 
using field theoretical 
methods\cite{boulat}, DMRG\cite{boulat,bohr} and other 
renormalization group approach\cite{borda,karrasch,andergassen}, and found the same qualitative 
behavior.

We consider the IRLM described by the 
Hamiltonian (in standard notation)
\beq
H&=&-\sum_{\a} i\a v \int dx \, \psi^{\dagger}_{\a}(x) 
\partial_{x}\psi_{\a}(x) +\varepsilon_{d}n_{d}   \nonumber\\
&+& \int dx\, U(x) 
\rho(x) n_{d} + 
\sum_{\a}\left[T_{\a}^{*}\psi^{\dagger}_{\a}(0)d + \mathrm{h.c.}\right],
\label{model}
\eeq
with $\a= \pm 1$ for $R$ and $L$ electrons, $n_{d}=d^{\dagger}d $
and $\rho = \sum_{\a} \rho_{\a} =\sum_{\a} \psi^{\dagger}_{\a} \psi_{\a}$. 
The dot-lead interaction $U(x)$ in Eq. (\ref{model})
can be either SR or LR.
The system is driven out of equilibrium by 
the bias perturbation $H_{B}=\sum_{\a}V_{\a}\int dx 
\rho_{\a}(x)=\sum_{\a}V_{\a}N_{\a}$ (with $N_{\a}$ 
the number of electrons with chirality $\a$).
For a non-perturbative treatement of the interaction 
we bosonize the fermion operators\cite{giamarchi,gonzbook}
\be
\psi_{\a}(x)=\frac{\eta_{\a}}{\sqrt{2\pi a} }
e^{-2\sqrt{\pi}\,i\a\phi_{\a}(x)},
\label{bospsi}
\ee
with boson field 
$\phi_{\a}(x)= i \a \sum_{q>0}\L_{q}
(b^{\dagger}_{\a q }e^{-i\a qx}-\mathrm{h.c.})
- \frac{\sqrt{\pi}xN_{\a}}{ \mathcal{L}}$ and 
$\eta_{\a}$ an anticommuting Klein factor. In the mode expansion of 
the boson field $\L_{q}=\frac{ e^{- \frac{aq}{2}} 
}{\sqrt{2\mathcal{L}q}}$, with $\mathcal{L}$ the length of the system 
and $a$ a short-distance cutoff.
The bosonized form of the  electron density takes the form 
$\rho_{\a}(x)=-\partial_{x}\phi_{\a}(x)/\sqrt{\pi}=\sum_{q>0}
\L_{q}q(
b^{\dagger}_{\a q }e^{-i\a qx}-\mathrm{h.c.}  ) 
+ \frac{N_{\a}}{ \mathcal{L}}$, and hence the bosonized Hamiltonian 
reads
\beq
H&=&\sum_{\a q} v q b^{\dagger}_{\a q} b_{\a q} +\varepsilon_{d}n_{d} 
\nonumber\\
&-& 
\sum_{\a q} \frac{\L_{q}q}{\sqrt{\pi}}\,
U_{q}(b^{\dagger}_{\a q} +b_{\a q}) n_{d} +  U_{0}\sum_{\a}
\frac{N_{\a}}{ \mathcal{L}} n_{d} \nonumber \\
&+&
\sum_{\a}
\left[
\frac{T_{\a}^{*}\eta^{\dagger}_{\a}}{\sqrt{2\pi a} }
e^{-2\sqrt{\pi}\sum_{q>0}\L_{q}
(b^{\dagger}_{\a q}- b_{\a q})} d +  \mathrm{h.c.}
\right],
\label{hboson}
\eeq
where 
$U_{q}=\int dx\, e^{iqx}U(x)$ and we used $U(x)=U(-x)$.
Next we perform a Lang-Firsov transformation to (formally) eliminate 
the dot-lead coupling. The unitary operator
$\mathcal{U}=e^{2\sqrt{\pi}\sum_{\a q} 
\frac{\L_{q}}{2\pi v}U_{q}
(b^{\dagger}_{\a q }-b_{\a q}  ) n_{d}}$
transforms the original Hamiltonian into $\tilde{H}=\mathcal{U}^{\dagger}H 
\mathcal{U}$ with
\beq
\tilde{H}&=&\sum_{\a q} v q b^{\dagger}_{\a q} b_{\a q} 
+\tilde{\varepsilon}_{d}n_{d} +  U_{0}\sum_{\a}
\frac{N_{\a}}{ \mathcal{L}} n_{d}  
\nonumber \\
&+&
\sum_{\a}
\left[T_{\a}^{*} f^{\dag}_{\a 0} d +  \mathrm{h.c.} \right] \, , \quad
\eeq
(from now on the sum will always be over $q>0$).
In the transformed Hamiltonian it appears the renormalized fermion field  
\be
f_{\a x}=\frac{\eta_{\a}}{\sqrt{2\pi a} }
e^{ 2\sqrt{\pi}\sum_{\b q}\L_{q} W_{\a \b q }
(b^{\dagger}_{\b q} e^{-i\a qx}- b_{\b q} e^{i\a qx})},
\ee
evaluated in $x=0$, with 
the effective interactions
$W_{RRq}=W_{LLq}=1+U_{q}/(2\pi v )$ and 
$W_{RLq}=W_{LRq}=U_{q}/(2\pi v )$, and the renormalized  energy  level
$\tilde{\ve}_{d}=\ve_{d}+\sum_{q}\frac{e^{-aq}}{\pi v 
\mathcal{L}}|U_{q}|^{2}$.
In the new basis the ground state of the 
isolated leads (i.e. for $T_{\a}=0$) is the vacuum $|0\rangle$
of the boson operators $b_{\a q}$. 
We can exploit this property to build the proper initial conditions by time 
propagation. We will consider the system initially 
uncontacted ($T_{\a}=0$), then switch on the contacts 
at time $t=0$ and let the current and dot-density relax. 
After relaxation, say at time $t_{0}$, we will bias the leads and 
study the screening dynamics from the transient to the steady state.
This procedure simulates with high accuracy the so-called 
{\it partition-free scheme}\cite{cini,stef}, as demonstrated in Refs. 
\cite{mssvl.2008,psc.2010,mcp.2011}.



We define the dot Green's function on the Keldysh contour as
\be
G(z,z')=\frac{1}{i} \bra {\cal T}\left\{d(z) d^{\dag}(z')\right\} \ket  ,
\ee
where ${\cal T}$ is the contour ordering, operators are in the 
Heisenberg picture with respect to 
$\tilde{H}+H_{B}$ (the bias perturbation does not change after the 
transformation), and  
the average is taken over 
the uncontacted ground state $|0\rangle\otimes |n \rangle$, 
$|n\rangle$ being the state of the dot 
with single ($n=1$) or zero ($n=0$) occupancy. The Green's function 
obeys the equation of motion (EOM)
\be
(i\partial_{z} - \tilde{\ve}_{d}) G(z,z')
= \d(z,z')+\sum_{\a} T_{\a}(z) G_{\a 0}(z,z'),
\label{eomg}
\ee 
where
$
G_{\a x} (z,z')=\frac{1}{i}\bra {\cal T}\left\{f_{\a x}(z) 
d^{\dag}(z') \right\}
\ket 
$ 
is the dot-lead Green's function\cite{note}. To close the EOM we 
derive  $G_{\a x}$ with respect to its first argument and find
\beq
&&\left(i\partial_{z}+i\a v \partial_{x}-V_{\a}(z)\right) G_{\a x} (z,z') 
\nonumber \\
&&=\frac{1}{i} \sum_{\b}\bra {\cal T}\left\{ \left[  T^{*}_{\b} f_{\b 0 }^{\dag} d
+ \mathrm{h.c.} 
, f_{\a x } 
\right](z) d^{\dag} (z)\right\} \ket .\quad \, 
\label{eom2}
\eeq
The computation of the correlator in the r.h.s. of Eq. (\ref{eom2}) is 
a formidable task. In order to proceed we approximate it by 
 $T^{*}_{\a} \bra \left( f_{\a 0 }^{\dag}  f_{\a x }+f_{\a x 
 }  f_{\a 0 }^{\dag}  \right)(z) \ket_{P} G(z,z') $, where $\bra \ldots \ket_{P}$ 
signifies that operators are in the Heisenberg picture with respect 
to the uncontacted Hamiltonian.
This approximation  is at the basis of our truncation scheme and 
becomes exact 
in the non-interacting case as well as in the uncontacted case.
Our approximation remains very accurate also for small 
$T_{\a}$ since it correctly reproduces recent results 
with SR dot-lead interaction (see below).

To solve the EOM for $G_{\a x}$ we define 
$g_{\a x \a x'}(z,z')=\frac{1}{i} \bra {\cal T}\left\{f_{\a x }(z)  
f_{\a x' }^{\dag}(z') \right\} \ket_{P}$ which satisfies the EOM
\beq
&&\left(i\partial_{z}+i\a v \partial_{x}-V_{\a}(z)\right) g_{\a x \a 
x'}(z,z')
\nonumber \\
&&=\d(z,z') \bra \left( f_{\a x }  f_{\a x' }^{\dag}+f_{\a x' 
 }^{\dag}  f_{\a x }  \right)(z) \ket_{P}.
\eeq
We can now perform a standard embedding and write the dot Green's 
function as the solution of
\be
(i\partial_{z}-\tilde{\ve}_{d})G(z,z') - \int_{\g} d\bar{z} \sum_{\a} 
\Sigma_{\a}(z,\bar{z}) G(\bar{z},z')=\d(z,z') ,
\label{eom4}
\ee
where
$
\Sigma_{\a}(z,z') =|T_{\a}|^{2}g_{\a 0 \a 0}(z,z') 
$
is the {\em correlated} embedding self-energy and the integral runs 
over the Keldysh contour.
Using the Langreth rules\cite{keldysh} 
Eq. (\ref{eom4}) is converted into a coupled system of Kadanoff-Baym
equations (KBE) which we solve numerically.
The real-time Keldysh components of $\Sigma$
can be evaluated exactly using the bosonization method\cite{giamarchi,gonzbook}
and read
\beq
\Sigma^{\lessgtr}_{\a}(t,t')=\pm \frac{i|T_{\a}|^{2}}{2\pi a } 
e^{-i\varphi_{\a}(t) } 
e^{Q[\pm (t-t')]}
e^{i\varphi_{\a}(t')  },
\label{sigmalutt}
\eeq
with phase $\varphi_{\a}(t)=\int^{t}_{0} d\bar{t}\, V_{\a}(\bar{t})$
and interaction dependent exponent
\be
Q(t)=\sum_{q}\frac{2 
\pi}{\mathcal{L}q}e^{-aq}(e^{ivqt}-1)\left[1-\frac{U_{q}}{\pi 
v}+\frac{1}{2}\left(\frac{U_{q}}{\pi v}\right)^{2}  \right].
\label{qfun}
\ee

From solution of Eq. (\ref{eom4}) we can easily calculate the 
dot-density from  $\bra 
n_{d}(t)\ket=-iG^{<}(t,t)$. 
\begin{figure}[tbp]
\includegraphics[width=8cm]{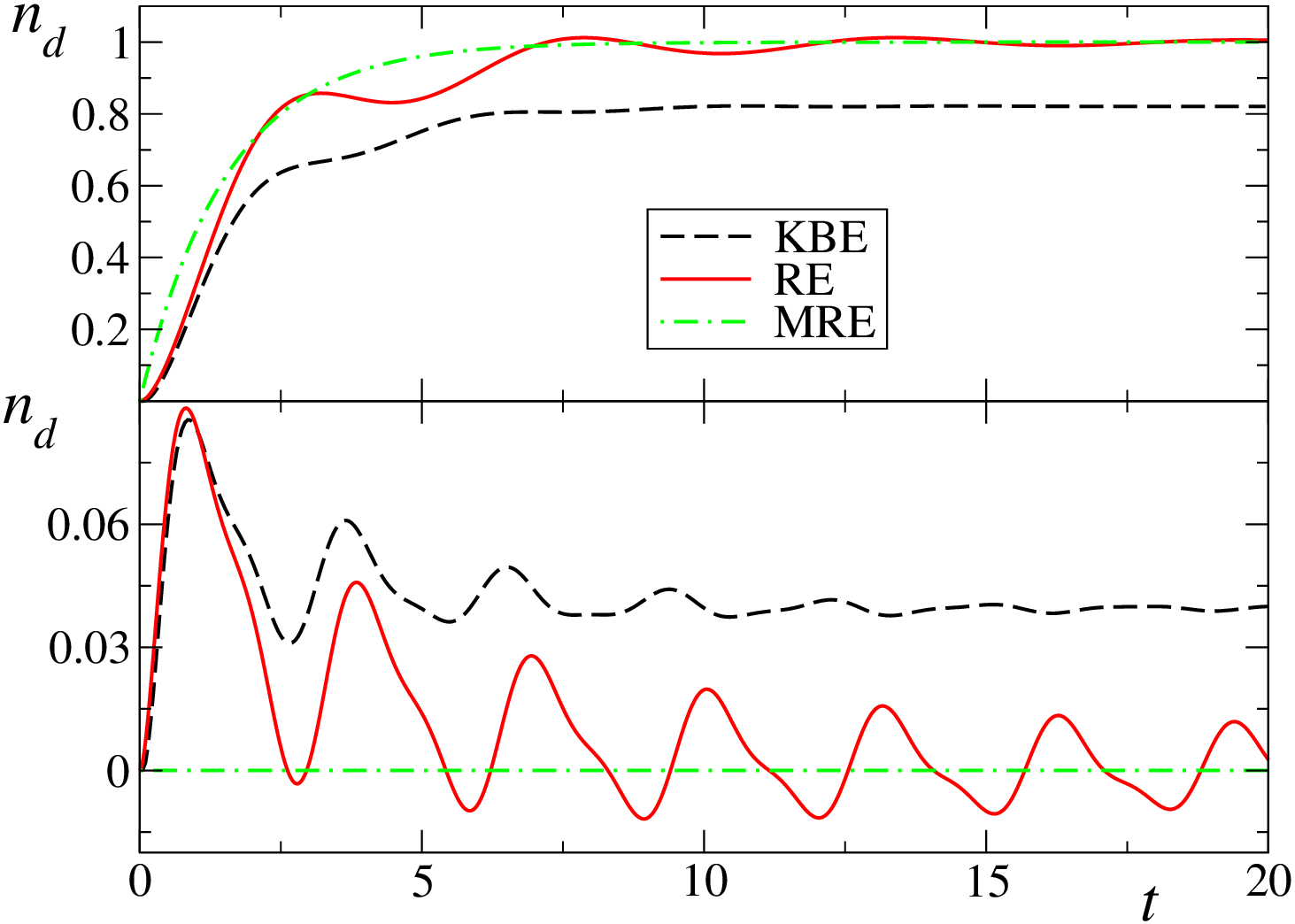}
\caption{Time-dependent density for SR intraction using KBE
(dashed), RE (solid) and MRE (dotted) for 
dot-energy $\tilde{\varepsilon}_{d}=0$ (top panel) and  
$\tilde{\varepsilon}_{d}=3$ (bottom panel) and initial dot-density $n=0$. The remaining 
parameters are $V_{L}=V_{R}=1$, 
$\gamma=0.2$, $U=0.5$, $t_{0}=0$ 
(partitioned scheme). Units are $10^{-1} v/a$ for $V_{\alpha}$, 
$\gamma $, $\tilde{\varepsilon}_{d}$, $10  a/v$ for $t$
and $ 2\pi v$ for $U$.}
\label{tddens}
\end{figure}
Similarly, the current $I_{\a}$ at the interface between the dot and 
lead $\a$ can be calculate from
\beq
I_{\a}(z)=\partial_{z} N_{\a}(z)&=& -i\, T^{*}_{\a} \bra  f^{\dag}_{\a 0}(z) d(z)   \ket
+\textrm{h.c.} \nonumber \\
&=&\int_{\g}d\bar{z} \,\Sigma_{\a}(z,\bar{z}) G(\bar{z},z) 
+\mathrm{h.c.}\quad
\label{tdcurr}
\eeq
In the steady-state regime $G(t,t')$ depends only on the time 
difference $t-t'$ and the current 
$\bar{I}=I_{L}(t\to\infty)=-I_{R}(t\to\infty)$ is given by 
a Meir-Wingreen-like formula
%
\beq
\bar{I}=
\int
\frac{d\omega}{2\pi} \frac{\Sigma^{>}_{L}(\omega) \Sigma^{<}_{R}(\omega) -
\Sigma^{<}_{L}(\omega) \Sigma^{>}_{R}(\omega) }
{|\omega - \tilde{\ve}_{d} -\sum_{\a}\Sigma_{\a}^{R}(\omega)|^{2}}.
\label{meir}
\eeq
Remarkably, the current cannot be written in terms of 
the difference between the leads Fermi functions despite the 
left and right contacts are the same. 

Our analysis starts by comparing the present 
approximation to the RE method,
recently employed in a similar context\cite{millis}.
In Fig. \ref{tddens} we plot the time-dependent dot-density 
using the KBE, the RE and their Markovian version (MRE)
for a SR interaction $U_{q}=U$. Both the KBE and RE densities exhibit 
oscillations with frequencies 
associated to charge-neutral excitations. As   
anticipated, however, the RE suffer from the negative-density 
problem\cite{whitney} (bottom panel). The MRE density is instead 
always non-negative but 
the lack of memory washes out the oscillations and the 
transient becomes a featureless exponential (top panel).
The KBE density is superior also at the steady state. Both the RE and 
MRE predict a zero-temperature steady-state density either 0 or 1 and hence severely 
underestimate the dot polarizability.
We also verified that the KBE approach is
charge-conserving since fulfills with high numerical 
accuracy the continuity 
equation $d \bra n_{d}(t) \ket /dt=I_{R}(t)+I_{L}(t)$
at every time (not shown).
%

\begin{figure}[t]
\includegraphics[width=7.2cm]{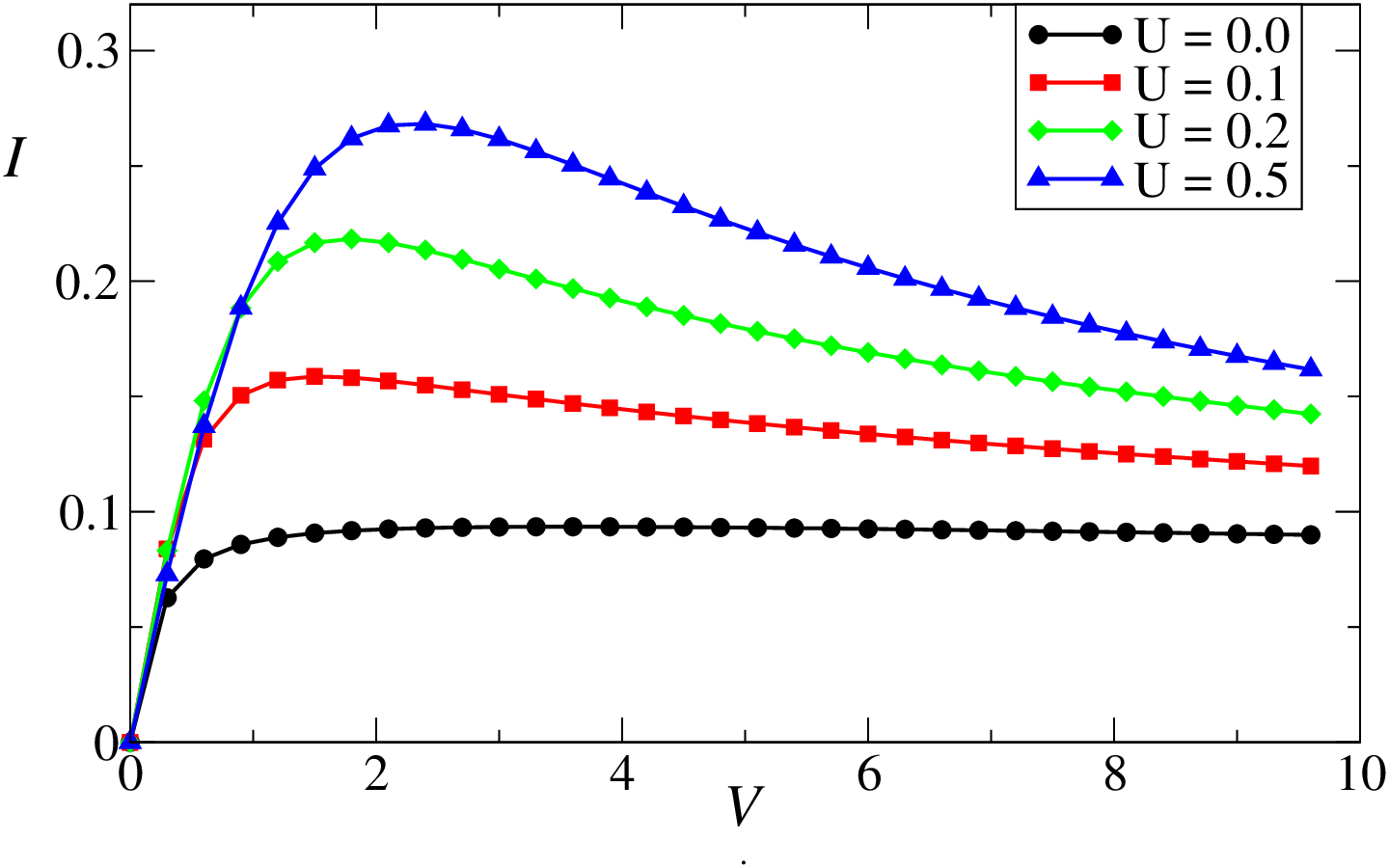}
\caption{$I$-$V$ curve for SR  
interaction. The parameters are 
$\gamma=0.2$, $\tilde{\varepsilon}_{d}=0$. Units are $10^{-2} v/a$ for $I$, $V$, 
$\gamma $, $\tilde{\varepsilon}_{d}$, and $2\pi v$ for $U$.}
\label{ivshort}
\end{figure}

We next calculate the steady-state current for a SR interaction and 
in the symmetric case ($\tilde{\ve}_{d}=0$, $T_{L}=T_{R}=T$, 
$V_{L}=-V_{R}=V>0$) recently considered by several 
authors\cite{boulat,karrasch,andergassen}.  
In this case the integral in Eq. (\ref{meir})
can be performed analytically and, by defining the tunneling rate 
$\gamma=|T|^{2}/v$, we have
\be
\bar{I}_{\mathrm{SR}}(V) \simeq
\left( 
\frac{Va}{v}\right)^{\beta-1}
\frac{ \gamma}{\pi \Gamma(\beta)} 
\tan^{-1}\left[
\frac{V}{\gamma}
\left( 
\frac{Va}{v}\right)^{1-\beta}
\Gamma(\b)
\right],
\label{ivsr}
\ee
with exponent $\b(U)=1+\frac{U(U-2\pi v)}{2\pi^{2}v^{2}}$. We notice 
that the functional form of $\bar{I}_{\mathrm{SR}}(V) $  is similar to the one derived 
in Ref. \cite{karrasch} within functional RG, although in the 
present case $\b$ is evaluated in a nonperturbative way.
The above expression (plotted in  Fig. \ref{ivshort}) is also in excellent 
agreement with the exact results  of 
Ref. \cite{boulat}.
In particular it reproduces the universal ohmic behavior 
$\bar{I}_{\mathrm{SR}}(V)\simeq V/\pi$ at small 
bias\cite{andrei} (with $\sigma_{0}=1/\pi$ the quantum of conductance), and the non-universal
power-law decay $\bar{I}_{\mathrm{SR}}(V)\sim 
V^{\beta -1}$  at large bias (the RE fail again here).
The Authors of Ref. \cite{boulat} observed numerically that
the $\beta$-exponent does not vary monotonically with $U$, and for a 
special value $\bar{U}$ of the interaction 
reaches the maximum value $\bar{\b}=1/2$, for which the IRLM is exactly 
solvable.
Our formula, which is valid for all $U$, is in fair good agreement with 
this result, and yields $\bar{U}=\pi v$. Note also that the 
steady-state current 
is symmetric around $\bar{U}$ since $\b(\bar{U}-\delta 
U)=\beta(\bar{U}+\delta U)$.

\begin{figure}[t]
\includegraphics[width=7.2cm]{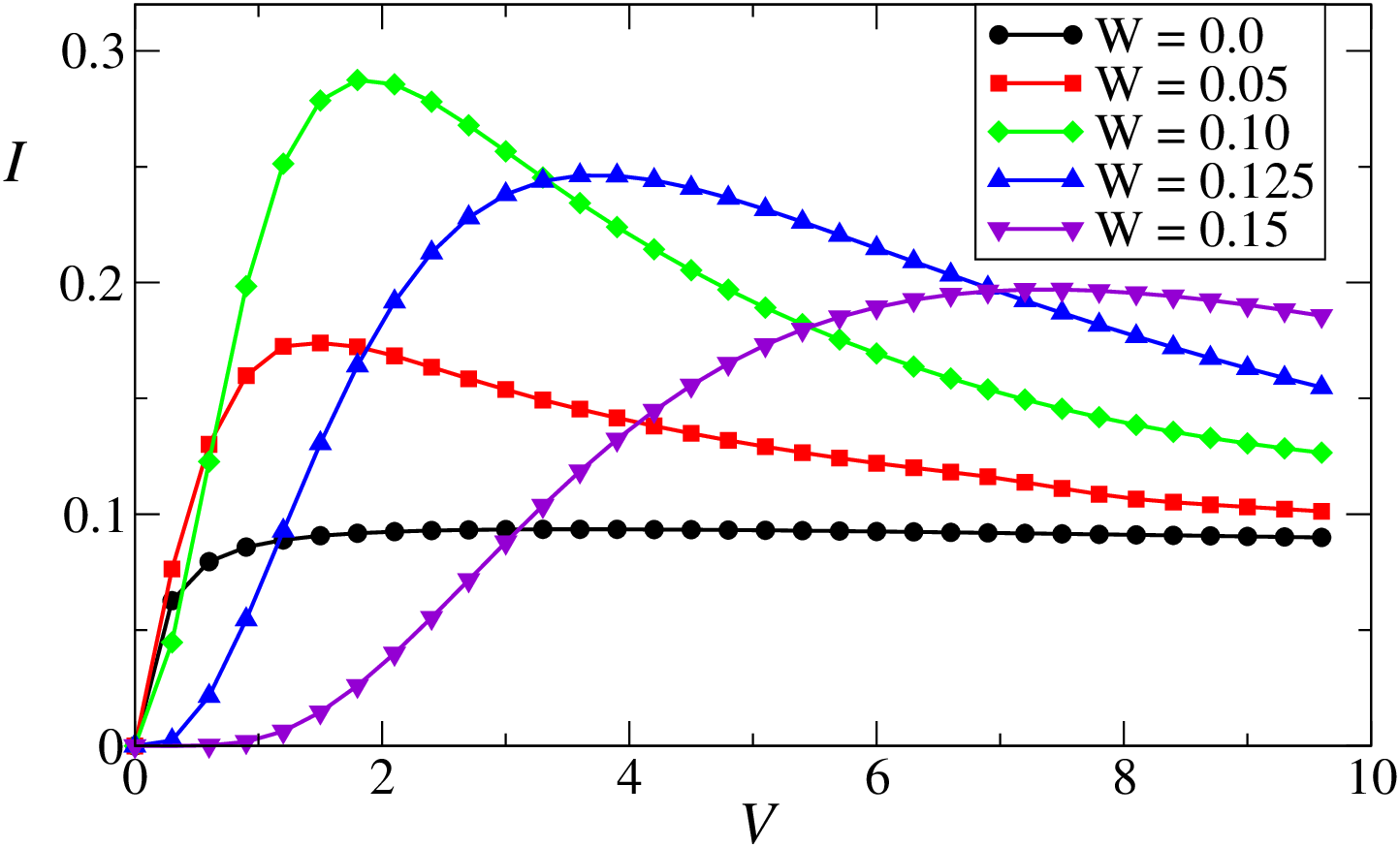}
\caption{$I$-$V$ curve for LR 
 interaction. Same parameters and units as in Fig. \ref{ivshort}.
 $W$ is in units of $2\pi v$.}
\label{ivlong}
\end{figure}

We can now present the most important numerical results of the paper, 
i.e., the time-dependent current with 
LR interaction $U_{q}=-W\ln 
(aq)^{2}$. In this case the function $Q(t)$ as well as
the integral in Eq. (\ref{meir}) must be evaluated numerically. 
In Fig. \ref{ivlong} we display the $I$-$V$ curve for several $W$'s.
The behavior is {\it qualitatively} different from the SR case.
In particular the zero-bias conductance is strongly suppressed with increasing $W$.
Due to the LR nature of the interaction the addition/removal of an 
electron to/from the dot induces a charge depletion/accumulation which 
extends smoothly deep inside the leads (jamming effect). 
For a current to flow the bias must be larger than the  
polarization energy of this particle-hole collective state. 
This picture also explains a common feature of the SR and LR $I$-$V$ 
curves, i.e., the existence of an optimal value of 
the interaction strenght for which the current has a maximum 
at fixed bias.
Increasing the interaction from zero the electron 
density diminishes close to the dot, thus enhancing the effective tunneling rate 
(Coulomb deblocking). However, increasing the interaction 
further the particle-hole binding energy becomes larger than the 
charge-transfer energy $V_{L}-V_{R}$ to move an electron from one lead to the 
other, and the current start decreasing. 
%
%

\begin{figure}[tbp]
\includegraphics[width=7.2cm]{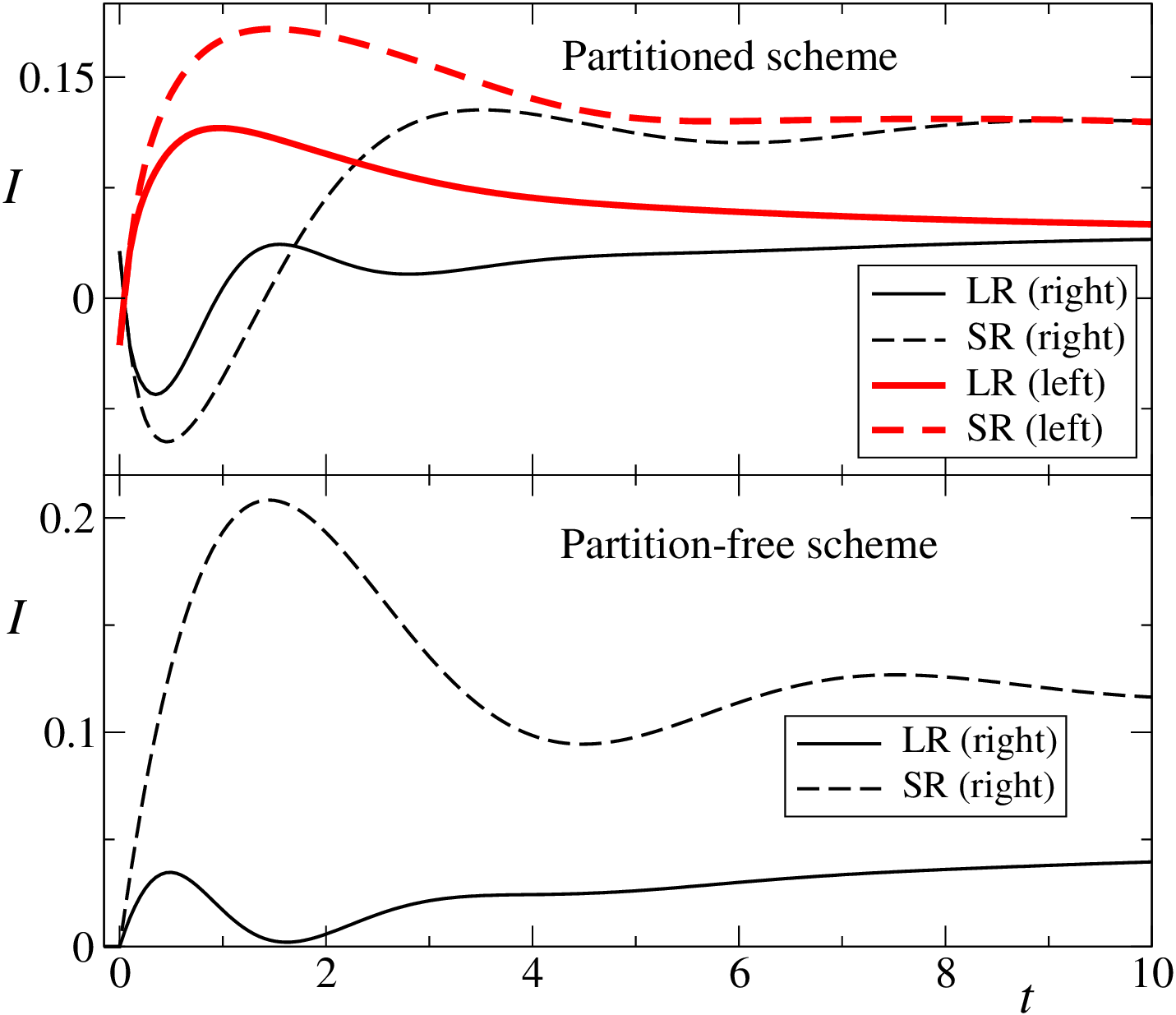}
\caption{Time-dependent current for SR (dashed) and LR (solid) 
interaction for an initial density $n=0$ . The thick curves represent 
$I_{L}(t)$ while the thin 
curves $I_{R}(t)$. The parameters are $V_{L}=-V_{R}=1$, 
$\gamma=0.2$, $U=W=0.2$, 
$\bar{\varepsilon}_{d}=0$. In the top panel $t_{0}=0$ (partitioned scheme) while 
in the bottom panel $t_{0}=60$ (simulated partition-free scheme).
 Units are $10^{-1} v/a$ for $V_{\alpha}$, 
$\gamma $, $\tilde{\varepsilon}_{d}$, $10  a/v$ for $t$
and $2\pi v$ for $U$ and $W$.
}
\label{tdplot}
\end{figure}

LR interactions have an impact also in the screening time.
In Fig. \ref{tdplot} we plot the time-dependent 
currents for LR and SR interaction with same interaction strength  
$W=U$\cite{nota}. The LR current relaxes faster 
both in the partitioned scheme (contacts and bias switched on 
simultaneously at $t=0$) and partition-free scheme.
The same behavior is observed 
 for different values of $W$ (not shown). 
The jamming effect of LR interactions is at the origin of the  
faster screening time. Electrons deep inside the leads 
suddenly respond to a change in the dot population induced by the 
applied bias. Finally we observe that the steady-state value of the 
current is the same in both schemes. This agrees with the results of 
Refs. \cite{stef,mssvl.2008,psc.2010,mcp.2011} according to which the memory 
of the initial state is washed out in the long-time limit.


In conclusion we presented a comprehensive characterization 
of the transport properties of the IRLM with LR 
interaction. We proposed an embedding scheme based on a 
suitable truncation of the EOM for the dressed fermion fields
and derived KBE which we solved numerically and benchmarked against 
available exact results.  
The method was compared with recently proposed 
RE approaches, and found to be superior from the transient (no 
negative densities) to the 
steady-state regime (no severe underestimation 
of the dot polarizability).
LR interactions leave
clear fingerprints in the time-dependent current as well as in the $I$-$V$ 
curve, and we believe that these features should survive when more 
sophisticated junctions (interacting multi-level resonant models) 
are considered.



\begin{thebibliography}{99}

%
    
\bibitem{book}
D. V. Averin and K. K Likharev,
{\it Mesoscopic Phenomena
in Solids},
(Elsevier, Amsterdam, 1991).


\bibitem{kondo}
A. C. Hewson, 
{\it The Kondo Problem to Heavy Fermions}
(Cambridge University Press, Cambridge, 1993).

\bibitem{cb}
H. Grabert and M. H. Devoret, 
{\it Single Charge Tunneling: Coulomb Blockade Phenomena in
Nanostructures},
(Plenum, New York, 1992).


\bibitem{thygesen}
K. S. Thygesen and A. Rubio,
Phys. Rev. Lett. \textbf{102}, 046802 (2009).

\bibitem{thygesen2}
M. Strange, C. Rostgaard, H. HŠkkinen, and K. S. Thygesen,
Phys. Rev. B {\bf 83}, 115108 (2011)

\bibitem{bescond}
Changsheng Li, M. Bescond, and M. Lannoo,
Phys. Rev. B \textbf{80}, 195318 (2009).

\bibitem{boulat}
E. Boulat, H. Saleur, and P. Schmitteckert,
Phys. Rev. Lett. \textbf{101}, 14061 (2008).

\bibitem{nishino}
A. Nishino, T. Imamura, and N. Hatano,
Phys. Rev. Lett. {\bf 102}, 146803 (2009).

\bibitem{karrasch}
C. Karrasch, M. Pletyukhov, L. Borda, and V. Meden,
Phys. Rev. B {\bf 81}, 125122 (2010).

\bibitem{andergassen}
S. Andergassen, M. Pletyukhov, D. Schuricht, H. Schoeller, and L. 
Borda,
Phys. Rev. B \textbf{81}, 205103 (2010).

\bibitem{borda}
L. Borda, K. Vlad\'ar, and A. Zawadowski, 
Phys. Rev. B {\bf 70}, 125107 (2007).

\bibitem{bohr}
D. Bohr and P. Schmitteckert,
Phys. Rev. B {\bf 75}, 241103R (2007).

\bibitem{meir}
Y. Meir and N. S. Wingreen,
Phys. Rev. Lett. \textbf{68}, 2512 (1992).

\bibitem{jauho}
A.-P. Jauho, N. S. Wingreen, and Y. Meir,
Phys. Rev. B \textbf{50}, 5528 (1994).

\bibitem{millis}
F. Elste, D. R. Reichman, and A. J. Millis,
Phys. Rev. B  \textbf{83}, 245405 (2011).

\bibitem{whitney}
R. S. Whitney,
J. Phys. A: Math. Theor. \textbf{41}, 175304 (2008).


\bibitem{giamarchi}
T. Giamarchi, {\em Quantum Physics in One Dimension} 
(Clarendon, Oxford, 2004).

\bibitem{gonzbook}
J. Gonz\`{a}lez, M. A. Mart\'{i}n-Delgado, G. Sierra and M. A. H. Vozmediano, 
\textit{Quantum Electron Liquids and High-$T_c$ Superconductivity}, 
Springer-Verlag, Berlin (1995). 


\bibitem{cini}
M. Cini,
Phys. Rev. B \textbf{22}, 5887 (1980).

\bibitem{stef}
G. Stefanucci and C. O. Almbladh,
Phys. Rev. B \textbf{69}, 195318 (2004).

\bibitem{mssvl.2008}
P. My\"oh\"anen, A. Stan, G. Stefanucci, and R. van Leeuwen,
Europhys. Lett. {\bf 84}, 67001 (2008).

\bibitem{psc.2010}
E. Perfetto, G. Stefanucci and M. Cini, 
Phys. Rev. Lett. {\bf 105}, 156802 (2010).

\bibitem{mcp.2011}
V. Moldoveanu, H. D. Cornean and C.-A. Pillet, 
Phys. Rev. B {\bf 84}, 075464 (2011).

\bibitem{note}
The term $ 
\frac{U_{0}}{\mathcal{L}} \frac{1}{i} \bra 
{\cal T}\left\{ N_{\a} (z) d(z) d^{\dag}(z')\right\}\ket
$
renormalizes the energy $\tilde{\ve}_{d}$. 
Indeed 
$N_{\a}/\mathcal{L}|0\ket=\rho_{\a}|0\ket$ with $\rho_{\a}$ the 
equilibrium density  of lead 
$\a$. Since the dot induces fluctuations of order $O(1/\mathcal{L})$ 
we have $\frac{U_{0}}{\mathcal{L}}\frac{1}{i} \bra 
{\cal T}\left\{N_{\a} (z) d(z) d^{\dag}(z')\right\}\ket = U_{0}\rho_{\a} 
G(z,z') +O(1/\mathcal{L})$.

\bibitem{keldysh}
R. van Leeuwen, N. E. Dahlen, G. Stefanucci, C. O. 
Almbladh, and U. von Barth, {\it Time-Dependent Density Functional Theory}
(Springer, New York, 2006);
Lect. Notes Phys. {\bf 706}, 33 (2006);
M. Cini, {\it Topics and Methods in Condensed Matter Theory}
(Springer-Verlag, Berlin, 2007).

\bibitem{andrei}
P. Mehta and N. Andrei, 
Phys. Rev. Lett. {\bf 96}, 216802 (2006).

\bibitem{nota}
This comparison is based on the idea of having an interaction 
$U_{q}=U[(1-\lambda)-\lambda \ln 
(aq)^{2}]$ parametrically dependent on $\lambda$. 
$U_{q}$ is SR for $\lambda=0$ whereas is LR for $\lambda=1$.































\end{thebibliography}
\end{document}